\documentclass{IEEEoj}

\usepackage{cite}
\usepackage{amsmath,amssymb,amsfonts}
\usepackage{algorithmic}
\usepackage{graphicx,color}
\usepackage{textcomp}
\def\BibTeX{{\rm B\kern-.05em{\sc i\kern-.025em b}\kern-.08em
    T\kern-.1667em\lower.7ex\hbox{E}\kern-.125emX}}
\AtBeginDocument{\definecolor{ojcolor}{cmyk}{0.93,0.59,0.15,0.02}}
\def\OJlogo{\vspace{-4pt}\hskip-4pt\includegraphics[height=18pt]{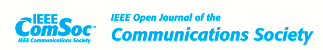}}
\usepackage{graphicx}
\usepackage{amsmath, amssymb}

\ifCLASSINFOpdf
\usepackage{float, multirow}
\else
\usepackage{cite}
\fi
\usepackage{xcolor}
\usepackage{cite}
\usepackage{hyperref}
\usepackage{balance}
\hypersetup{%
  citebordercolor=blue
}
\begin{document}
\bstctlcite{IEEEexample:BSTcontrol}

\title{Noise-Domain Non-Orthogonal Multiple Access}

\author{Erkin Yapici\IEEEauthorrefmark{1} \IEEEmembership{(Graduate Member, IEEE)}, Yusuf Islam Tek \IEEEauthorrefmark{1,2} \IEEEmembership{(Graduate Member, IEEE)}, AND Ertugrul Basar\IEEEauthorrefmark{1,3}
\IEEEmembership{(Fellow, IEEE)}}
\affil{Communications Research and
Innovation Laboratory (CoreLab), Department of Electrical and Electronics
Engineering, Koç University, Sariyer 34450, Istanbul, Turkiye}
\affil{Turk Telekom R\&D Department, Altindag, 06080, Ankara, Turkiye}
\affil{Department of Electrical Engineering, Tampere University, 33720 Tampere, Finland}
\corresp{CORRESPONDING AUTHOR: Ertugrul Basar (e-mail: ertugrul.basar@tuni.fi and ebasar@ku.edu.tr).}
\authornote{This work is supported by the Scientific and Technological Research Council of Türkiye (TÜBİTAK) through the 1515 Frontier R\&D Laboratories Support Program for the Türk Telekom 6G R\&D Lab (Project No. 5249902) and under Grant No. 124E146.}
\markboth{Noise-Domain Non-Orthogonal Multiple Access}{Yapici \textit{et al.}}

\begin{abstract}
In this paper, we present noise-domain non-orthogonal multiple access (ND-NOMA), an innovative communication scheme that utilizes the modulation of artificial noise mean and variance to convey information. Distinct from traditional methods such as power-domain non-orthogonal multiple access (PD-NOMA) that heavily rely on successive interference cancellation (SIC), ND-NOMA utilizes the noise domain, considerably reducing power consumption and system complexity. Inspired by noise modulation, ND-NOMA provides lower bit error probability (BEP), making it highly suitable for next-generation Internet-of-things (IoT) networks. Our theoretical analyses and computer simulations reveal that ND-NOMA can achieve exceptionally low bit error rates in both uplink and downlink scenarios, in the presence of Rician fading channels. The proposed multi-user system is supported by a minimum distance detector for mean detection and a threshold-based detector for variance detection, ensuring robust communication in low-power environments. By leveraging the inherent properties of noise, ND-NOMA offers a promising platform for long-term deployments of low-cost and low-complexity devices.
\end{abstract}


\maketitle

\section{INTRODUCTION}
\IEEEPARstart{O}{rthogonal} multiple access (OMA) techniques offer advantages such as effective interference management, simplicity of implementation, and fair resource allocation. However, they also present several disadvantages, including reduced efficiency as the number of users increases, challenges in dynamic resource allocation, scalability issues in large-scale networks, and limited flexibility due to the need to maintain orthogonality among users \cite{8357810,8970580,9691334}. While OMA is well-suited for current wireless networks, future scenarios with high-density and dynamic demands may benefit from newer techniques like non-orthogonal multiple access (NOMA). 

Unlike traditional OMA techniques, where different users are allocated distinct time, frequency, or code resources, NOMA allows multiple users to share the same time and frequency resources simultaneously. This principle is achieved primarily through two main NOMA techniques: power-domain NOMA (PD-NOMA) and code-domain NOMA (CD-NOMA) \cite{7676258, 3214321321}. PD-NOMA uses different power levels to multiplex users, employing superposition coding at the transmitter and successive interference cancellation (SIC) at the receiver. SIC decodes the strongest signal first, iteratively subtracting it to decode subsequent signals. However, imperfect signal cancellation can hinder its effectiveness. CD-NOMA multiplexes users in the code domain using techniques like low-density spreading, sparse code multiple access, lattice-partition multiple access, multi-user shared access, and pattern-division multiple access. These techniques separate user signals with distinct code sequences, offering an alternative to PD-NOMA.

\begin{figure*}[t]
    \centering
    \begin{minipage}{0.49\textwidth}
        \centering
        \includegraphics[width=\textwidth]{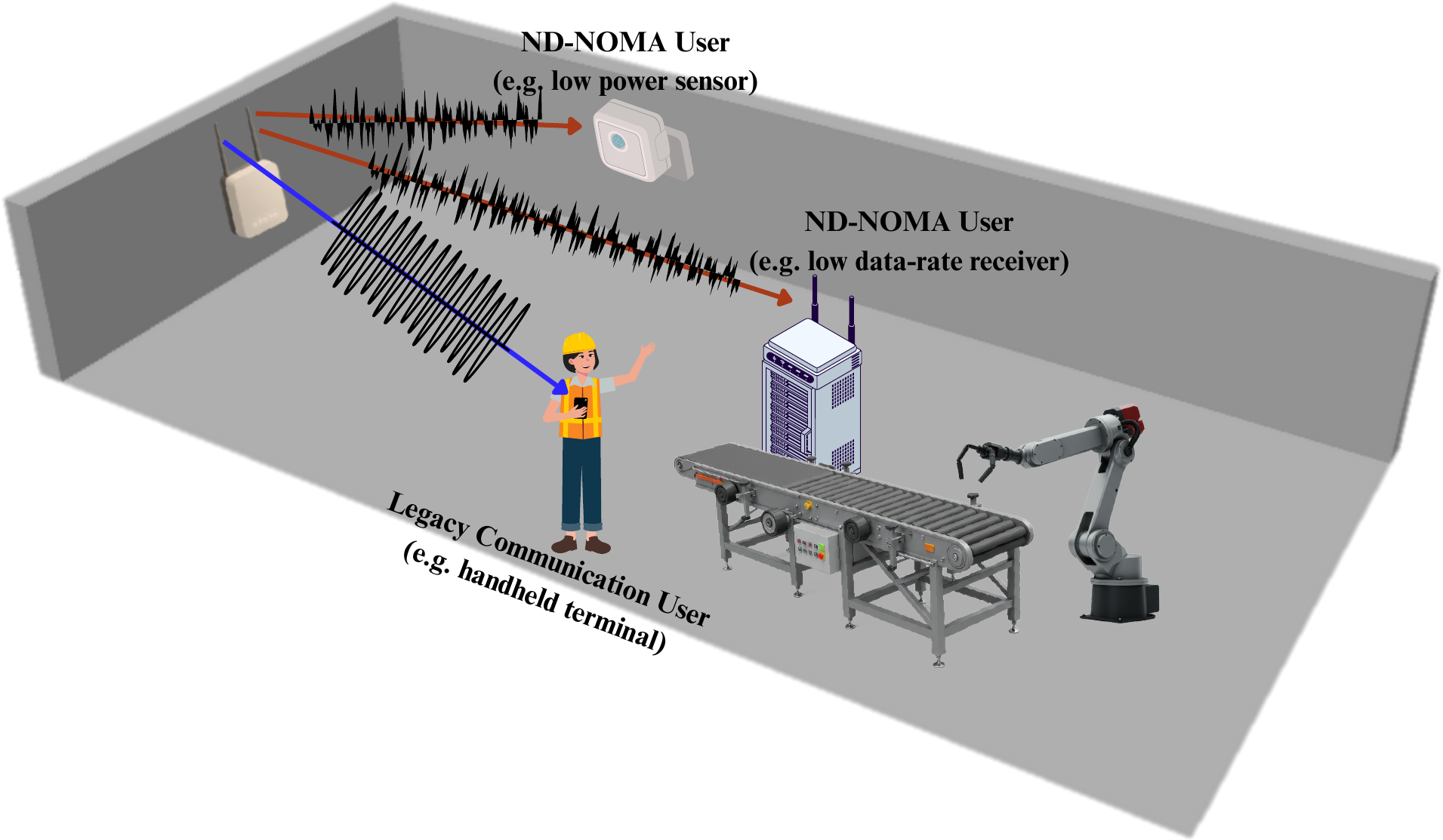}
        \\ \text{(a)}
        \label{fig:systemmodel_uplink}
    \end{minipage}%
    \hfill
    \begin{minipage}{0.49\textwidth}
        \centering
        \includegraphics[width=\textwidth]{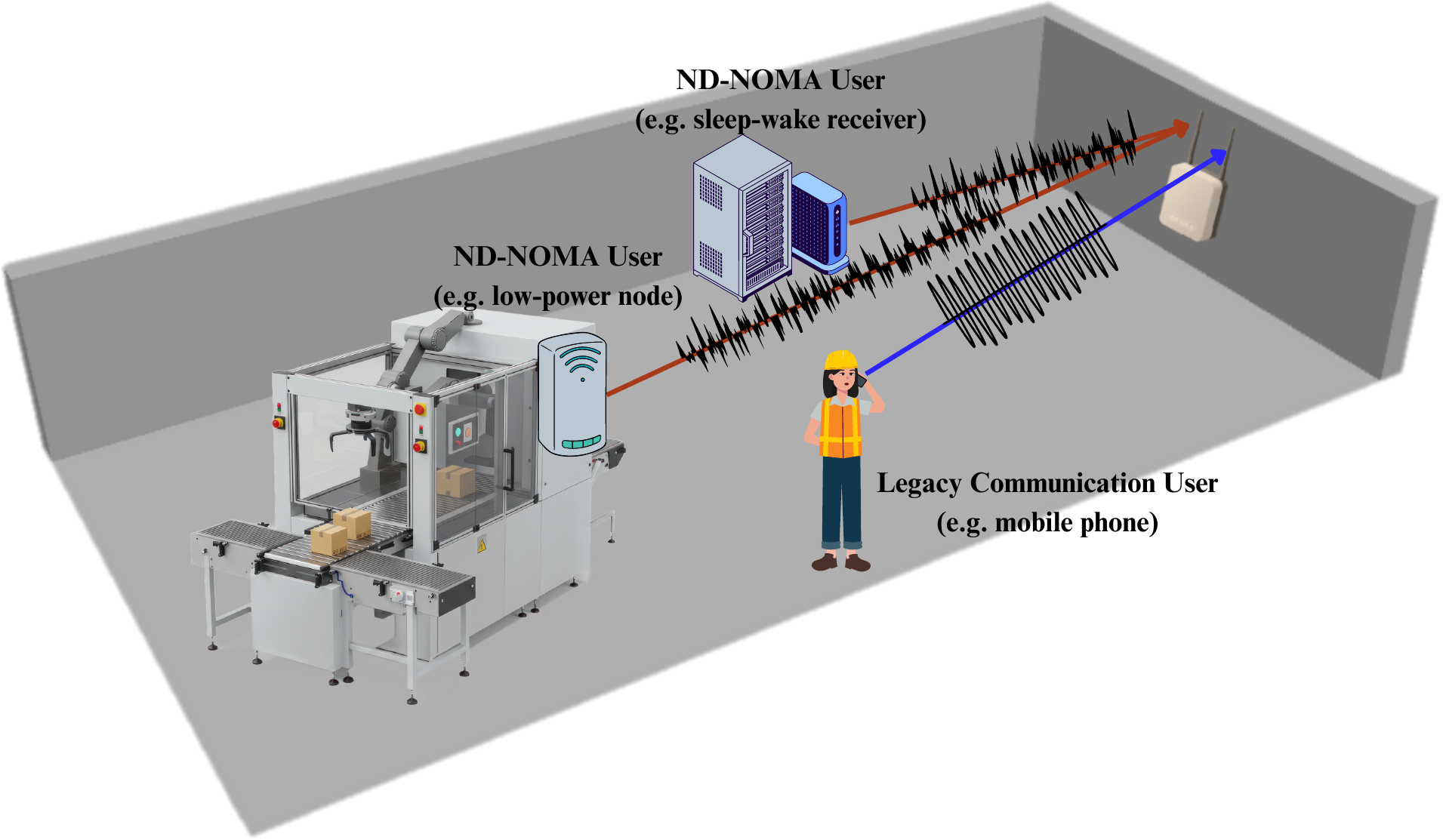}
        \\ \text{(b)}
        \label{fig:systemmodel_downlink}
    \end{minipage}
    \caption{{Illustration of a potential ND-NOMA use-case in low-data-rate, low-energy IoT environments, where (a) represents the downlink transmission and (b) represents the uplink transmission. For illustration, the legacy user is depicted as a sinusoidal carrier, whereas ND-NOMA users exhibit a noise-based waveform.}}
    \label{fig:system_model}
\end{figure*}


However, NOMA systems face several challenges, such as optimizing power allocation to balance energy efficiency and quality-of-service, requiring accurate channel state information (CSI) for effective SIC operation, and managing residual interference from imperfect SIC. Additionally, there are trade-offs between spectral and energy efficiency, receiver design complexities, SIC error propagation, and sensitivity to channel gain measurements \cite{8972353}. Recent studies have also explored enhancing the physical layer security of uplink NOMA systems through the use of energy-harvesting jammers, offering a promising direction for secure and energy-aware NOMA designs \cite{9193903}. Interference management, integration with carrier aggregation, and addressing security concerns are also critical for NOMA systems \cite{9943006}.

In earlier times, thermal noise modulation (TherMod) was introduced to convey information using thermal noise \cite{basar2022thercom}. This concept recently evolved into schemes that transmit data using parameters like random signal variance. A transceiver architecture later emerged, using either artificially generated or thermal noise sources \cite{10373568}. Similarly, noise loop modulation had been utilized to deliver artificial noise in a feedback loop between legitimate users, enhancing unconditional security by preventing eavesdropping without requiring knowledge of the eavesdropper's channel \cite{9740497}. Parallel to these studies, during transmission variance-based sampling was also previously used\cite{sigg2012calculation}. It has been shown that noise modulation might provide advantages for simplifying the receiver architectures and robustness to certain channel effects and system impairments. This motivates the consideration of its application in multi-user environments.

\begin{figure*}[!t]
    \centering
    \includegraphics[width=\linewidth]{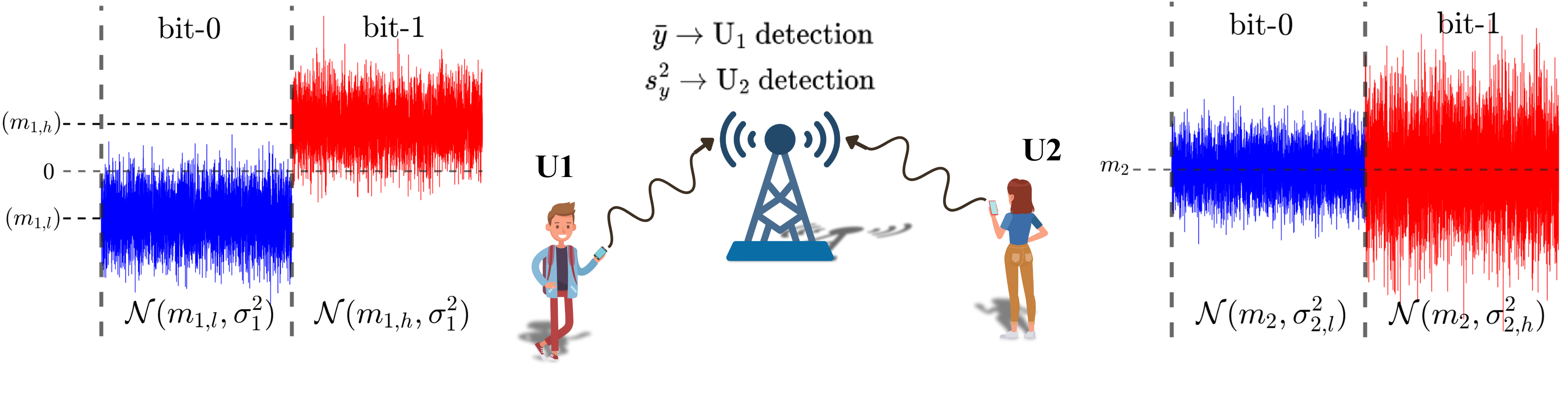}
    \caption{Uplink ND-NOMA scheme with two users using real Gaussian signals.}
    \label{fig:uplink}
\end{figure*}

In this work, by taking the noise modulation concept one step further, we propose an innovative communication scheme called noise-domain NOMA (ND-NOMA). The fascinating property of the Gaussian distribution is that the sum of two independent Gaussian variables remains Gaussian \cite{6868214}. In ND-NOMA, we exploit this by adding two Gaussian noise signals, each carrying different user information. The resulting signal retains its Gaussian nature, simplifying processing and detection at the receiver, which allows the ND-NOMA to efficiently handle multiple signals with minimal error, ideal for low-power IoT networks. This scheme utilizes the mean and variance of the artificial noise signals to transmit data. Basically, bits of one user are modulated using the mean of the noise samples, while bits of the other user are modulated using the variance of the noise samples.
Since ND-NOMA leverages Gaussian noise samples, the system does not require processing steps such as frequency synchronization for the downlink scenario.
Thus, ND-NOMA significantly reduces power consumption and system complexity, making it ideal for the low power consumption requirements of next-generation applications such as Internet-of-things (IoT) networks. Our theoretical analysis and computer simulations demonstrate that ND-NOMA can achieve exceptionally low bit error rate (BER) in both uplink and downlink scenarios. The robust performance of the system is supported by a minimum distance detector for mean detection and a threshold-based detector for variance detection, ensuring reliable communication even in low-power scenarios. Our findings indicate that ND-NOMA has excellent performance, particularly in Rician fading channels, and can be utilized for long-term and large-scale deployments of low-cost, low-complex devices.

In emerging IoT use-cases, legacy systems were not performing well enough in terms of scalability, so the low-data-rate communication schemes were essential. Unlike PD-NOMA or CD-NOMA, designed for high-throughput scenarios, ND-NOMA offers a noise-domain access strategy well suited for low-data-rate IoT environments. As next-generation multiple access technology is a key candidate for 6G, ND-NOMA aligns well with the access needs of future wireless communication scenarios. ND-NOMA is well-suited for integration with Device B types of ambient IoT systems, as highlighted in \cite{butt2023ambientiotmissinglink}. Device B, with its limited energy storage and semi-passive operation, aligns with ND-NOMA’s low-power and low-complexity demands, essential for sustainable IoT deployments. For Device Type B, the goal is to have low complexity, and with its simple receiver architecture, ND-NOMA is well-suited for low-data-rate scenarios (0.1–5 kbps). Device B is semi-passive, uses backscattering with limited energy storage, and has power consumption below 1 mW. ND-NOMA removes the need for SIC and strict frequency and phase synchronization, reducing computational and energy costs at receivers, and making it ideal for large-scale IoT, thereby further lowering complexity and power consumption for practical deployments.
 Studies in ISAC show that random waveforms enhance spectral efficiency, complementing ND-NOMA’s goals of low-power communication \cite{Koivunen_2024}. Moreover, energy harvesting techniques such as SWIPT further support ND-NOMA's energy-efficient design \cite{Song_2014, Ashraf_2021}. It is worth mentioning that the ND-NOMA scheme is not designed for broadband cellular communication scenarios, but instead it is designed for low-power and low-data-rate IoT environments, and our main motivation is to decompose the signals without using SIC. Using noise statistics such as mean and variance to convey information requires a sufficient number of noise samples per symbol for accurate estimation. Without increasing the symbol duration or the sampling frequency, estimation accuracy may degrade, particularly in fast-varying channels. Fig. \ref{fig:system_model} provides a generic representation of the ND-NOMA concept within the context of IoT applications; however, the specific operational mechanisms and signal structures are more clearly illustrated in Figs. \ref{fig:uplink} and \ref{fig:downlink}. ND-NOMA suits low-data-rate smart city applications such as environmental monitoring, smart metering, and public infrastructure telemetry, where energy efficiency, scalability, and asynchronous transmission are essential at the receivers.

This article is structured as follows. Section II covers uplink ND-NOMA theory and BEP optimization, Section III addresses downlink performance, Section IV presents numerical results, and Section V concludes the paper.

\section{Uplink ND-NOMA: System Model and Performance Analysis}
In this section, we first provide the system model for uplink ND-NOMA for two users and then introduce a general framework to assess the theoretical BEP performance of the two users.

\subsection{System Model}
As shown in Fig. 2, in the uplink transmission of two users, one of the users (User 1, $\text{U}_1$) uses the mean of the transmitted Gaussian samples. In contrast, the other user (User 2, $\text{U}_2$) exploits the variance as in noise modulation schemes. Specifically, denoting the $n$th noise samples of $\text{U}_1$ and $\text{U}_2$ respectively by $s_1^n$ and $s_2^n$, for $\text{U}_1$, the transmission of information bits is accomplished by alternating the mean of the samples between low and high values, that is, for bit-\(0\) and bit-\(1\), we have, $s_1^n \sim \mathcal{N}(m_{1,l},\sigma_1^2)$ and $s_2^n \sim \mathcal{N}(m_{1,h},\sigma_1^2)$. As discussed later, we set $m_{1,h}=-m_{1,l}$ for optimum performance. We assume that two users consider $N$ noise samples for each bit.

On the other hand, for $\text{U}_2$, we apply noise modulation by alternating the noise variance only, and for bit-\(0\) and bit-\(1\), we have $s_2^n \sim \mathcal{N}(m_2,\sigma_{2,l}^2)$ and $s_2^n \sim \mathcal{N}(m_2,\sigma_{2,h}^2)$. At this point, we set the mean of samples to zero, that is, $m_2=0$, to minimize the interference to $\text{U}_1$. This initial model assumes binary-level signaling, while a generalization is subject to further studies.

In this setup, we assume an average transmission power of \(P\) for each user, which corresponds to the second moment of the transmitted samples\footnote{Assuming equiprobable \(\text{U}_2\) bits, over a long sequence, half of its samples will have \(\sigma_{2,l}^2\) variance while the other half having \(\sigma_{2,h}^2\). As a result, the second moment of \(\text{U}_2\) samples converges to \(P\).}: $E[(s_{1}^{n})^{2}] = E[(s_{2}^{n})^{2}]=P$, which equals the sum of the squared mean and variance of their samples. In light of this, $P=m_{1,l}^2 + \sigma_1^2$ for $\text{U}_1$. Here, $\text{U}_1$ dedicates $\beta$ portion of its available power for the variance of its samples while dedicating a $(1-\beta)$ portion for the mean, that is, $\sigma_1^2=\beta P$ and $m_{1,l}^{2} =  m_{1,h}^{2}=(1-\beta)P $. Since $m_2=0$, all power of $\text{U}_2$ is dedicated to the variance of its samples, $P=(\sigma_{2,l}^2+\sigma_{2,h}^2)/2$.

In what follows, we use the terminology considered in the study of \cite{basar2022thercom} and define $\delta=\sigma_{2,l}^2/\sigma_{w}^2$ as the ratio of useful and disruptive noise variances. Additionally, \(\alpha\) stands for the ratio of high and low variance values for $\text{U}_2$, that is, $\sigma_{2,h}^2=\alpha \sigma_{2,l}^2$. Accordingly, we obtain  $\eta=\sigma_{1}^2/\sigma_{w}^2=(1+\alpha)\delta \beta/2$. 

In light of the model in Fig. 2, the received $n$th noise sample at the receiver is expressed as
\begin{equation}
y^n = h_1 s_{1}^{n} + h_2 s_{2}^{n} + w^n, \quad n=1,\ldots,N.
\label{eq:system_model}
\end{equation}
Here, $h_1$ and $h_2$ are the complex baseband channel coefficients of the links between $\text{U}_1$-base station (BS) and $\text{U}_2$-BS, respectively, and $w^n\sim \mathcal{CN}(0,\sigma_{w}^2)$ is the complex baseband additive white Gaussian noise sample at the BS. Our analysis and computer simulations assume either unit-gain Rayleigh or Rician fading models. Accordingly, for Rayleigh fading, we have $h_1,h_2 \sim \mathcal{CN}(0,1)$ while for Rician fading with $K$ parameter, we have $h_{1,R},h_{1,I},h_{2,R},h_{2,I} \sim \mathcal{N}\left(\sqrt{\frac{K}{2(1+K)}},\frac{1}{2(1+K)}\right)$. Then, conditioned on channel coefficients, we obtain
\begin{align}
    \mathrm{E}[y^n]=& h_1 m_{1,i}  \nonumber \\
    \mathrm{VAR}[y^n]=& |h_1|^2 \sigma_1^2  + |h_2|^2 \sigma_{2,k}^2 + \sigma_w^2,\quad i,k\in \{l,h\}.
\end{align}

The task of the BS is to process the received samples of \eqref{eq:system_model} through statistical tests to detect the information bits of both users. The following two subsections present minimum distance-based detection rules for both users and derive their theoretical bit error probability (BEP) performance.

\subsection{Uplink - User 1 Detection}
The task of the BS during the detection of $\text{U}_1$'s bit involves calculating the sample mean of the received samples, \(y^n\), $n=1,\ldots,N$, followed by a decision process. In light of this, the sample mean is calculated as
\begin{equation}
\bar{y} = \frac{1}{N} \sum\limits_{n=1}^{N} y^n,
\label{eq:sample_mean}
\end{equation}
which is complex Gaussian distributed with mean $\mathrm{E}[\bar{y}]=h_1 m_{1,i}$ (unbiased estimate) and $\mathrm{VAR}[\bar{y}]=(|h_1|^2 \sigma_1^2  + |h_2|^2 \sigma_{2,k}^2 + \sigma_w^2)/N$ variance with $ i,k\in \{l,h\} $. Based on this, the following binary hypothesis test is considered at the BS to detect $\text{U}_1$'s bit:
\begin{equation}
\hat{b}_1=\begin{cases}
	0,\quad \text{if}\quad \left|  \bar{y}-h_1 m_{1,l} \right|^2 < \left|  \bar{y}-h_1 m_{1,h} \right|^2   \\
	1,\quad \text{if}\quad \left|  \bar{y}-h_1 m_{1,h} \right|^2 < \left|  \bar{y}-h_1 m_{1,l} \right|^2.
\end{cases}
\end{equation}
The above detector is a ``minimum distance detector" and considering the complex Gaussian distribution of $\bar{y}$, it is a \textit{maximum likelihood} one. In light of this and also considering the symmetry for bit-$0$ and bit-$1$, conditioned on channel coefficients, the BEP of $\text{U}_1$ becomes
\begin{equation} 
P_b =  P\left( \left. |\bar{y} - h_1 m_{1,h}|^2 < |\bar{y} - h_1 m_{1,l} |^2 \right| b_1=0 \right).
\label{eq:5}
\end{equation}
Expanding the terms in \eqref{eq:5} and considering $m_{1,h}=-m_{1,l}$, we obtain
\begin{equation}
	P_b=P\left( 4\mathrm{Re} \left\lbrace \bar{y} h_1^* m_{1,l}  \right\rbrace<0  \left. \right| b_1=0 \right). 
\end{equation}
Defining the Gaussian distributed random variable $D=\mathrm{Re} \left\lbrace \bar{y} h_1^* m_{1,l}  \right\rbrace$, we obtain, $P_b=P(D<0\left. \right| b_1=0 )=Q\left(  m_D/\sigma_D \right)  $, where $\mathrm{E}[D]=m_D$ and $\mathrm{VAR}[D]=\sigma_D^2$ are conditional statistics of $D$. While $m_D$ can be easily obtained as $m_D=\left|h_1 \right|^2 m_{1,l}^2 $, derivation of $\sigma_D^2$ is not straightforward due to correlation of $\bar{y}_R$ and $\bar{y}_I$. Specifically, we obtain
\begin{align}
	\sigma_D^2 &=\text{VAR} \left[ \mathrm{Re} \left\lbrace (\bar{y}_R+j\bar{y}_I)(h_{1,R}+jh_{1,I}) m_{1,l} \right\rbrace  \right] \nonumber \\& = \text{VAR}\left[  m_{1,l} \left( \bar{y}_R h_{1,R}  + \bar{y}_I h_{1,I}\right)     \right] \nonumber \\
	&=m_{1,l}^2\left(h_{1,R}^2 \mathrm{VAR}\left[ \bar{y}_R \right] + h_{1,I}^2 \mathrm{VAR}\left[ \bar{y}_I \right]  \right. \nonumber \\
	& \quad\quad\quad\quad\quad + \left.  2 h_{1,R}h_{1,I} \mathrm{COV}(\bar{y}_R,\bar{y}_I)  \right). 
	\label{eq:7}
\end{align}
Substituting the following variance values in \eqref{eq:7} that are obtained after tedious calculations,
\begin{align}
\mathrm{VAR}\left[ \bar{y}_R \right]=& \frac{1}{N} \left( h_{1,R}^2 \sigma_1^2 + h_{2,R}^2 \sigma_{2,k}^2 + \sigma_{w}^2 /2 \right) \nonumber \\
\mathrm{VAR}\left[ \bar{y}_I \right]=& \frac{1}{N} \left( h_{1,I}^2 \sigma_1^2 + h_{2,I}^2 \sigma_{2,k}^2 + \sigma_{w}^2 /2 \right) \nonumber \\
\mathrm{COV}(\bar{y}_R,\bar{y}_I) = & \frac{1}{N} \left( h_{1,R}  h_{1,I}  \sigma_1^2 + h_{2,R}  h_{2,I}  \sigma_{2,k}^2  \right), 
\end{align}
BEP of  $\text{U}_1$ can be obtained as $P_b=Q\left(  m_D/\sigma_D \right)$, which is conditioned on $h_1$, $h_2$, and $\sigma_{2,k}^2$. Due to the complexity of the terms inside the $Q$ function, we resort to numerical integration methods over the probabilistic distributions of $h_1$ and $h_2$ to derive the unconditional BEP. Specifically, we obtain
\begin{align}
&\bar{P}_b=\!\!\int\!\!\!\int\!\!\!\int\!\!\!\int\! P_b f(h_{1,R}) f(h_{1,I}) f(h_{2,R}) f(h_{2,I})  \nonumber\\
&\hspace{4cm} d h_{1,R} d h_{1,I}d h_{2,R} d h_{2,I},
\label{eq:9}
\end{align}
where a further averaging is needed over the two equiprobable values of $\sigma_{2,k}^2$ for $k\in \left\lbrace l,h \right\rbrace $. In order to calculate the unconditional BEP, we encountered a highly complex integral that cannot be solved analytically. Consequently, we utilized numerical methods, specifically Monte Carlo integration, to perform the evaluation. The detailed methodology for this approach is provided in Appendix A and is used to calculate the unconditional BEP among the users on both downlink and uplink scenarios.

Finally, we note that, as NoiseMod schemes, conditional BEP is a decaying function of $\sqrt{N}$ (through $\sigma_D^2$).

\subsection{Uplink - User 2 Detection}
In contrast to $\text{U}_1$'s mean detection problem, the task of the BS is to formulate a variance detection problem to extract $\text{U}_2$'s bits. While we work on a real and positive search space for this problem, complex distributions of $y^n$ and $\bar{y}$, more importantly, their correlated real and imaginary components, pose a severe challenge to the BEP derivation of $\text{U}_2$.

In light of this, the sample variance of the received samples is calculated as
\begin{equation}
s_y^2 = \frac{1}{N-1} \sum\limits_{n=1}^{N} \left| y^n - \bar{y} \right|^2.
\label{eq:sample_variance}
\end{equation}
Considering the two possible variance values of $y^n$, given as $ s_0^2=\sigma_1^2 |h_1|^2 + \sigma_{2,l}^2 |h_2|^2  + \sigma_w^2$ and $s_1^2=\sigma_1^2 |h_1|^2 + \sigma_{2,h}^2 |h_2|^2  + \sigma_w^2 $, further we obtain,
\begin{align}
	s_0^2 &= \sigma_w^2 \left(1 + \eta\left| h_1 \right|^2 +  \delta \left| h_2 \right|^2  \right) \nonumber \\
	s_1^2 &= \sigma_w^2 \left(1 + \eta\left| h_1 \right|^2 +  \alpha\delta \left| h_2 \right|^2  \right).
\end{align}
Here, $\eta$, $\delta$, and $\alpha$ are system-related constants defined in Section II.A. Defining a variance threshold as $\gamma=\chi \sigma_w^2$, where $\chi$ is the scaled threshold value \cite{10373568}, and it is defined as 
\begin{equation}
    \chi = \frac{2(1+\delta)(1+\alpha\delta)}{2+\delta(1+\alpha)}.
\end{equation}
Thus, $\text{U}_2$ detection problem is formulated as
\begin{equation}
	\hat{b}_2=\begin{cases}
		0,\quad \text{if}\quad s_y^2 < \gamma  \\
		1,\quad \text{if}\quad s_y^2 > \gamma,
	\end{cases}
\end{equation}

\begin{figure*}[!t]
    \centering
    \includegraphics[width=0.90\textwidth]{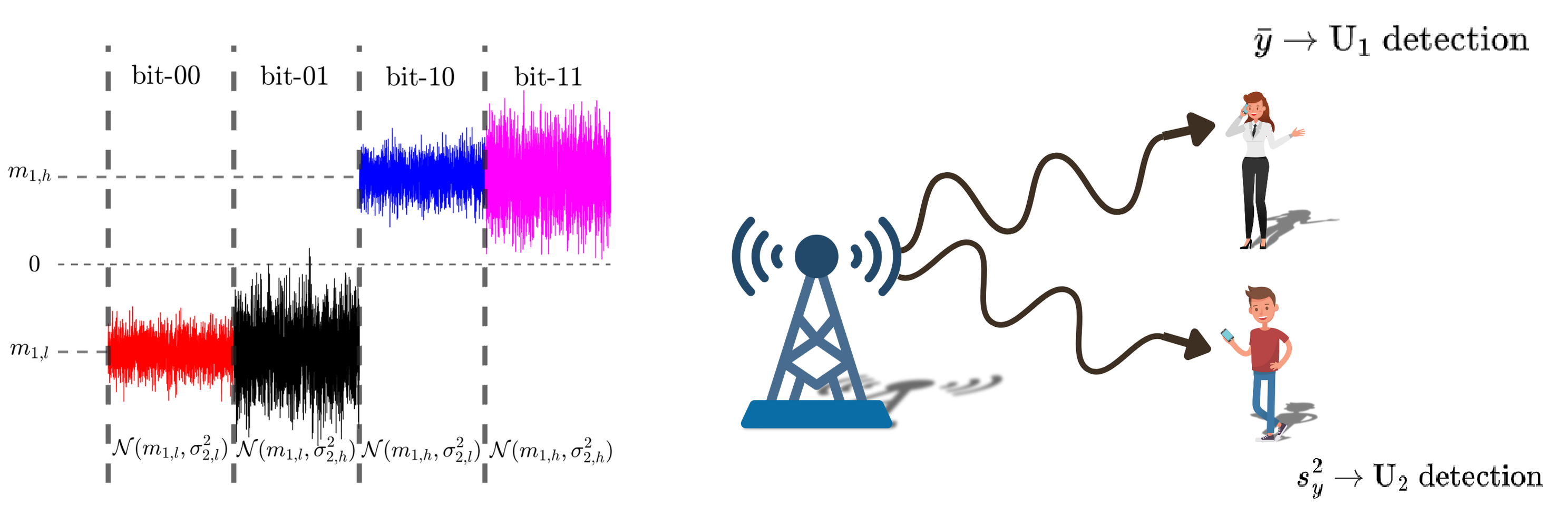}
    \caption{Downlink ND-NOMA scheme with two users using real Gaussian signals.}
    \label{fig:downlink}
\end{figure*}

Accordingly, BEP of  $\text{U}_2$ bits is obtained as
\begin{equation} 
	P_b =  \frac{1}{2}P\left( \left. s_y^2 >\gamma  \right| b_2=0 \right) + \frac{1}{2}P\left( \left. s_y^2 <\gamma \right|  b_2=1 \right).
	\label{eq:13}
\end{equation}
This calculation requires the statistics of $ s_y^2  $, which will be derived next.

At this point, for large $N$, to simply the theoretical derivations due to correlated components, we assume $s_y^2 \approx \frac{1}{N-1} \sum\nolimits_{n=1}^{N} \left| y^n - h_1 m_{1,i} \right|^2$, where $\bar{y}$ is replaced by its mean $\mathrm{E}[\bar{y}]=h_1 m_{1,i}$, $i\in \left\lbrace l,h\right\rbrace $. Accordingly, \eqref{eq:sample_variance} can be expressed in the quadratic form of $2N$ real Gaussian random variables:
\begin{align}
	s_y^2& = \frac{1}{N-1} \sum\limits_{n=1}^{N} \left( y^n_R-h_{1,R} m_{1,i}  \right)^2 +  \left( y^n_I-h_{1,I} m_{1,i}  \right)^2 \nonumber \\
	&= \mathbf{y}^{\text{T}} \Lambda\mathbf{y},
	\label{eq:quadratic}
\end{align}
where $\mathbf{y} \sim \mathcal{N} \left(\mathbf{0}, \Sigma \right) $ is a $2 N \times 1$ vector with Gaussian distributed elements and $\Lambda$ is a $2 N \times 2N$ diagonal matrix. Here, $\Sigma$ is a banded covariance matrix, given as
\begin{equation}
	\Sigma=\begin{bmatrix}
		\sigma_R^2 & c & 0 & \ldots & &  0 \\
		c & \sigma_I^2 & 0 & \ldots & & 0 \\
		0 & 0 & \ddots & & &  0 \\
		\vdots & \vdots & \ldots  & 0 & \sigma_R^2 & c \\
		 0 & 0 & \ldots  & 0 & c & \sigma_I^2  
			\end{bmatrix}_{2N\times 2N}
			\label{eq:15}
\end{equation}
where $\sigma_R^2=\mathrm{VAR}[y_R^n]=h_{1,R}^2 \sigma_1^2 + h_{2,R}^2 \sigma_{2,k}^2 + \sigma_w^2 /2$, $\sigma_I^2=\mathrm{VAR}[y_I^n]=h_{1,I}^2 \sigma_1^2 + h_{2,I}^2 \sigma_{2,k}^2 + \sigma_w^2 /2$, and $c=\mathrm{COV}( \left( y^n_R-h_{1,R} m_{1,i}  \right),\left( y^n_I-h_{1,I} m_{1,i}  \right) )$ for $i,k \in\left\lbrace l,h \right\rbrace $ and $n=1,\ldots,N$. After simple manipulations, we obtain $c=h_{1,R}h_{1,I} \sigma_1^2 + h_{2,R}h_{2,I} \sigma_{2,k}^2$.

At this point, interested readers might resort to advanced statistics by considering series expansions for the distribution of Gaussian quadratic forms or advanced mathematical packages on generalized chi-square distributions \cite{Mathai_1992}. To simplify our analysis, we consider the central limit theorem (CLT) and assume that $s_y^2 \sim \mathcal{N}(\mu_s,\sigma_s^2)$ for large enough $N$.

From \cite{Paolella_2019}, we obtain the mean and variance of $s_y^2$ as
\begin{align}
\mu_s &= \frac{1}{N-1} \mathrm{tr} [\Lambda \Sigma] = \frac{N(\sigma_R^2 + \sigma_I^2)}{N-1} \nonumber\\
\sigma_s^2 &= \frac{2}{(N-1)^2} \mathrm{tr}[\Lambda \Sigma \Lambda \Sigma]= \frac{2N}{(N-1)^2} \left( \sigma_R^4 + \sigma_I^4 + 2c^2 \right),  
\label{eq:16}
\end{align}
which are consistent with the definition of $s_y^2$ in \eqref{eq:sample_variance}. Accordingly, BEP is obtained as
\begin{equation}
P_b=\frac{1}{2}Q\left( \frac{\gamma-\mu_{s\left. \right|b_2=0 }}{\sigma_{s\left. \right|b_2=0}} \right) + \frac{1}{2} Q \left( \frac{\mu_{s\left. \right|b_2=1}-\gamma}{\sigma_{s\left. \right|b_2=1}}\right)  
\label{eq:17}
\end{equation}
where $\mu_s$ and $\sigma_s$ are conditioned on the transmitted bit through $\sigma_{2,k}^2$. Selecting the threshold for equal bit error probabilities of bit $0$ and bit $1$ yields
\begin{equation}
\gamma= \frac{\sigma_{s\left. \right|b_2=0}\times \mu_{s\left. \right|b_2=1 } + \sigma_{s\left. \right|b_2=1}\times \mu_{s\left. \right|b_2=0 }  }{\sigma_{s\left. \right|b_2=0}+ \sigma_{s\left. \right|b_2=1}}.
\label{eq:18}
\end{equation}
Substituting this value in \eqref{eq:17} yields the conditional BEP as
\begin{equation}
	P_b=Q\left( \frac{\mu_{s\left. \right|b_2=1}-\mu_{s\left. \right|b_2=0} } {\sigma_{s\left. \right|b_2=0}+ \sigma_{s\left. \right|b_2=1} } \right). 
\label{eq:19}	
\end{equation}
As $\text{U}_1$'s BEP, for $N \gg 1$, we observe that $\text{U}_2$'s BEP is a decaying function of $\sqrt{N}$ through the $\sigma_s^2$ term in the $Q$ function. 

Despite its simple detection architecture and our assumptions above, the final result for the BEP is still too complicated for taking an average over channel realizations, and we have to use numerical integration as in \eqref{eq:9} to obtain unconditional BEP. Nevertheless, the computational complexity of our approach remains significantly lower than traditional PAM demodulation methods with SIC, which require iterative interference cancellation and complex computations. In contrast, our detection methods rely solely on straightforward mean, variance, and threshold calculations, making them far more efficient and less computationally demanding.

The following steps outline the key aspects of the uplink system model, summarizing the transmission and detection processes:

\begin{itemize}
    \item \textit{Step 1}:  
    U$_1$ encodes bits by adjusting the mean of its Gaussian noise signal, while U$_2$ encodes bits by varying the variance with its mean kept at zero to avoid interference.  

    \item \textit{Step 2}:  
    The base station receives the combined signals from U$_1$ and U$_2$, affected by channel fading and noise.

    \item \textit{Step 3}:  
    The base station decodes U$_1$’s bit by computing the mean of received samples and uses a minimum distance detector, while U$_2$’s bit is decoded by computing the variance and applying a threshold-based detector.
\end{itemize}

\section{Downlink ND-NOMA: System Model and Performance Analysis}
In this section, we focus on downlink ND-NOMA for two users by defining its system model and then presenting its theoretical BEP performance.

\begin{table}[t!t!]
				\caption{Statistics of the transmitted and received signals for Downlink ND-NOMA}
	\begin{tabular}{c|c|c}
		$\text{U}_1/\text{U}_2$ bits & $s_{BS}^n  $     & $y_p^n,\, p \in \left\lbrace 1,2 \right\rbrace  $ \\  \hline 
		00         &  $\mathcal{N}(m_{1,l},\sigma_{2,l}^2)$  &  $\mathcal{CN}(h_p m_{1,l},\left| h_p\right|^2 \sigma_{2,l}^2+\sigma_w^2)$                                                \\
		01         &   $\mathcal{N}(m_{1,l},\sigma_{2,h}^2)$  &       $\mathcal{CN}(h_p m_{1,l},\left| h_p\right|^2 \sigma_{2,h}^2+\sigma_w^2)$                                            \\
		10         &   $\mathcal{N}(m_{1,h},\sigma_{2,l}^2)$    &       $\mathcal{CN}(h_p m_{1,h},\left| h_p\right|^2 \sigma_{2,l}^2+\sigma_w^2)$                                           \\
		11         &     $\mathcal{N}(m_{1,h},\sigma_{2,h}^2)$     &      $\mathcal{CN}(h_p m_{1,h},\left| h_p\right|^2 \sigma_{2,h}^2+\sigma_w^2)$                                           
	\end{tabular}
    \label{tab:downlink_NDNOMA}
\end{table}

\subsection{System Model}
As shown in Fig. \ref{fig:downlink}, for the downlink scheme, a composite signal is generated by the BS considering the information bits of two users. Here, we again consider binary modulation for simplicity. Accordingly, BS's transmitted noise samples follow $s_{BS}^n \sim \mathcal{N}(m_{1,i},\sigma_{2,k}^2)$ distribution for $ i,k \in \left\lbrace l,h \right\rbrace $. In other words, the mean of the transmitted signal is dictated by $\text{U}_1$'s bit while its variance is determined according to $\text{U}_2$'s bit. In this case, the received signals at two users are given by
\begin{align}
	y_1^n & = h_1 s_{BS}^n + w_1^n \nonumber \\
	y_2^n & = h_2 s_{BS}^n + w_2^n,
\end{align}
for $n=1,\ldots,N$. Here, $h_1$ and $h_2$ stand for the downlink channel fading coefficients between the BS and users, and $w_1^n$ and $w_2^n$ are AWGN samples at users with $\mathcal{CN}(0,\sigma_w^2)$ distribution. Specifically, Table \ref{tab:downlink_NDNOMA} lists the distribution of the transmitted and received signals at all terminals depending on four user bit combinations, $00,01,10$, and $11$, where the first and second bits respectively stand for $\text{U}_1$ and $\text{U}_2$ bits for a given bit duration.

Here, we assume that the total transmission power is fixed to $P$, that is,  $E[(s_{BS}^{n})^{2}] =P$. Similar to the $\beta $ parameter of the uplink scheme, we define a new parameter, $\psi$, as the portion of $\text{U}_1$'s allocated power, that is,  $m_{1,l}^{2} =  m_{1,h}^{2}=\psi P $ (dc power of the transmitted signal). Accordingly, for $\text{U}_2$, we have  $(\sigma_{2,l}^2+\sigma_{2,h}^2)/2=(1-\psi)P$ (ac power of the transmitted signal). We again consider $m_{1,h}=-m_{1,l}$ and $\sigma_{2,h}^2=\alpha \sigma_{2,l}^2$. 

Since the users' signals do not overlap as in the uplink scheme, the detection model of the downlink system is much simpler, and there is no interaction among user signals. Furthermore, there is no need for successive interference cancellation and error propagation issues as in the downlink PD-NOMA scheme. In what follows, we provide detector architectures for both users.

\subsection{Downlink - User 1 Detection}
The receiver architecture of the downlink scheme for $\text{U}_1$ is very similar to that of the uplink scheme, and the BEP can be evaluated by simple modifications. Specifically, $\text{U}_1$ calculates the sample mean of its received samples as $\bar{y}_1 = \frac{1}{N} \sum\nolimits_{n=1}^{N} y^n_1$, which is also Gaussian distributed with $\mathrm{E}[\bar{y}_1]=h_1 m_{1,i}$ and $\mathrm{VAR}[\bar{y}_1]=(|h_1|^2  \sigma_{2,k}^2 + \sigma_w^2)/N$  for $ i,k\in \{l,h\} $. Accordingly, the minimum distance detector is formulated as
\begin{equation}
	\hat{b}_1=\begin{cases}
		0,\quad \text{if}\quad \left|  \bar{y}_1-h_1 m_{1,l} \right|^2 < \left|  \bar{y}_1-h_1 m_{1,h} \right|^2   \\
		1,\quad \text{if}\quad \left|  \bar{y}_1-h_1 m_{1,h} \right|^2 < \left|  \bar{y}_1-h_1 m_{1,l} \right|^2.
	\end{cases}
\end{equation}
Similar to the uplink scheme's BEP in \eqref{eq:5}, BEP for this case is obtained as
\begin{align} 
	P_b &=  P\left( \left. |\bar{y}_1 - h_1 m_{1,h}|^2 < |\bar{y}_1 - h_1 m_{1,l} |^2 \right| b_1=0 \right). \nonumber \\
	&= P\left( \mathrm{Re} \left\lbrace \bar{y}_1 h_1^* m_{1,l}  \right\rbrace<0  \left. \right| b_1=0 \right)\nonumber \\
	&= P(D<0\left. \right| b_1=0 )=Q\left(  m_D/\sigma_D \right).
	\label{eq:22}
\end{align}

Here, Gaussian distributed variable $D$ has the following statistics: $m_D=\left|h_1 \right|^2 m_{1,l}^2 $ and $\sigma_D^2 =m_{1,l}^2(h_{1,R}^2 \mathrm{VAR}\left[ \bar{y}_{1,R} \right] + h_{1,I}^2 \mathrm{VAR}\left[ \bar{y}_{1,I} \right]  + 2 h_{1,R}h_{1,I} \mathrm{COV}(\bar{y}_{1,R},\bar{y}_{1,I}) )$. With simple manipulations, variance values are obtained as
\begin{align}
	\mathrm{VAR}\left[ \bar{y}_{1,R} \right]=& \frac{1}{N} \left( h_{1,R}^2  \sigma_{2,k}^2 + \sigma_{w}^2 /2 \right) \nonumber \\
	\mathrm{VAR}\left[ \bar{y}_{1,I} \right]=& \frac{1}{N} \left( h_{1,I}^2 \sigma_{2,k}^2 + \sigma_{w}^2 /2 \right) \nonumber \\
	\mathrm{COV}(\bar{y}_{1,R},\bar{y}_{1,I}) = & \frac{1}{N}  h_{1,R}  h_{1,I}  \sigma_{2,k}^2 .
\end{align}
Conditional BEP is obtained by substituting these values in $\sigma_D^2$ first and then considering \eqref{eq:22}.

\begin{figure*}[t]
    \centering
    \begin{minipage}{0.49\textwidth}
        \centering
        \includegraphics[width=\textwidth]{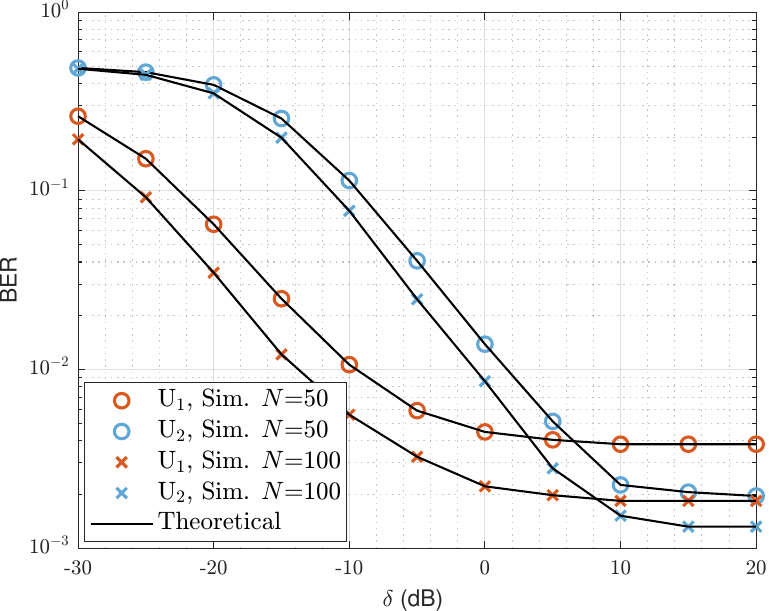}
        \\ \text{(a)}
        \label{fig:rayleigh_uplink}
    \end{minipage}%
    \hfill
    \begin{minipage}{0.49\textwidth}
        \centering
        \includegraphics[width=\textwidth]{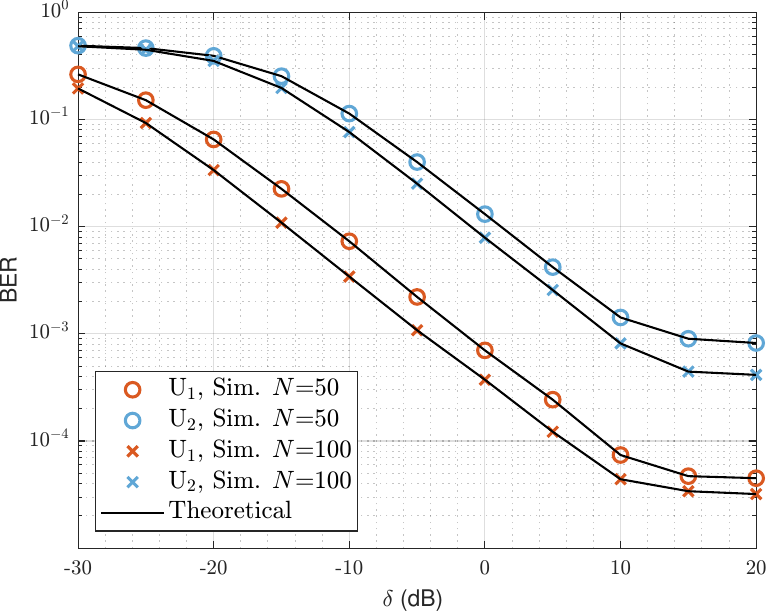}
        \\ \text{(b)}
        \label{fig:rayleigh_downlink}
    \end{minipage}
    \caption{Theoretical BEP and simulated BER for ND-NOMA versus $\delta = {\sigma_{2,l}^2}/{\sigma_w^2}$, in Rayleigh fading channels in (a) uplink and (b) downlink scenarios.}
    \label{fig:ALL_paper_rayleigh}
\end{figure*}

\subsection{Downlink - User 2 Detection}
Finally, this subsection investigates the receiver architecture and performance of $\text{U}_2$ for the downlink scheme. Fortunately, the same analytical framework of Section II.C can be considered here for slight modifications.

For $\text{U}_2$'s variance detection in the downlink scheme, the sample variance is obtained as
\begin{equation}
s_{y_2}^2 = \frac{1}{N-1} \sum\limits_{n=1}^{N} \left| y^n_2 - \bar{y}_2 \right|^2
\label{eq:sample_variance2}
\end{equation}
where $\bar{y}_2$ is the sample mean for the samples of $\text{U}_2$. Here, in a similar manner to uplink detection, we define a variance threshold as $\gamma=\chi \sigma_w^2$ and formulate the $\text{U}_2$ detection problem as
\begin{equation}
\hat{b}_2=\begin{cases}
0,\quad \text{if}\quad s_{y_2}^2 < \gamma  \\
1,\quad \text{if}\quad s_{y_2}^2 > \gamma.
\end{cases}
\end{equation}
In light of this, BEP of the $\text{U}_2$ bit is obtained as
\begin{equation} 
P_b =  \frac{1}{2}P\left( \left. s_{y_2}^2 >\gamma  \right| b_2=0 \right) + \frac{1}{2}P\left( \left. s_{y_2}^2 <\gamma \right|  b_2=1 \right).
\label{eq:26}
\end{equation}
\begin{figure*}[t]
    \centering
    \begin{minipage}{0.49\textwidth}
        \centering
        \includegraphics[width=\textwidth]{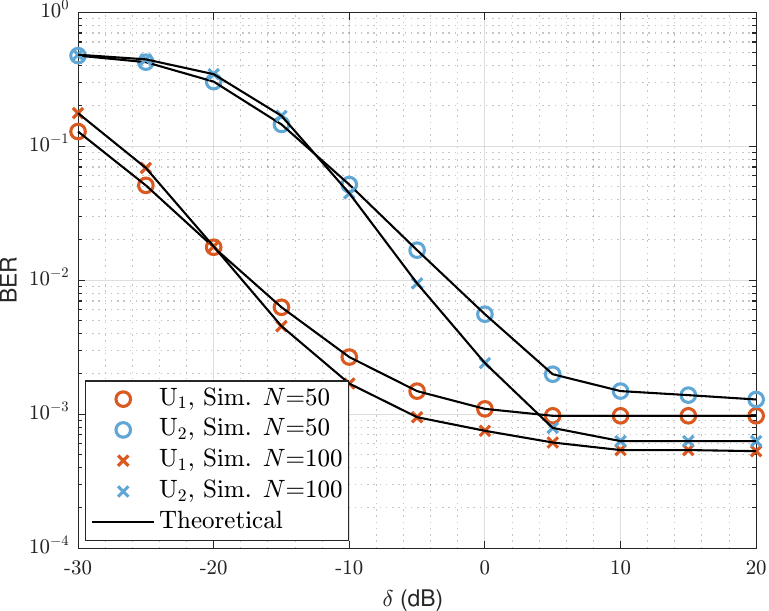}
        \\ \text{(a)}
        \label{fig:simulation_results_ul_k5}
    \end{minipage}%
    \hfill
    \begin{minipage}{0.49\textwidth}
        \centering
        \includegraphics[width=\textwidth]{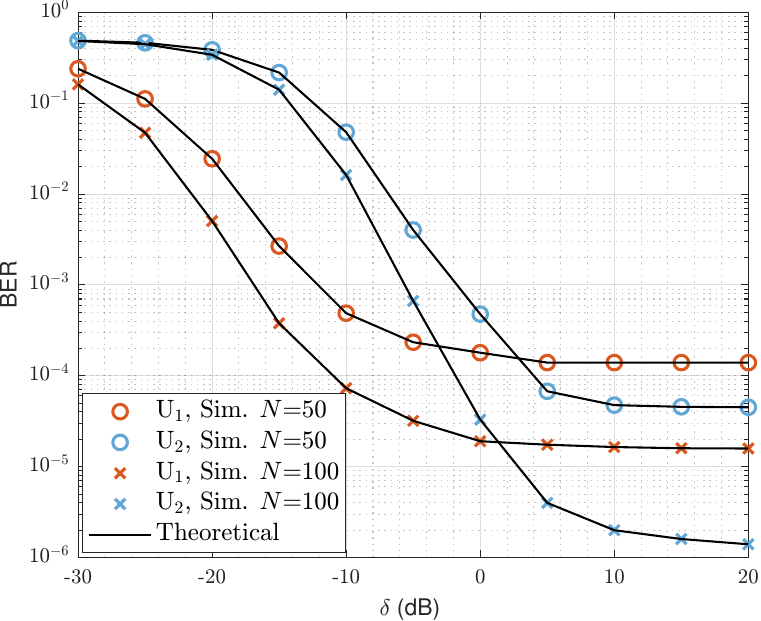}
        \\ \text{(b)}
        \label{fig:simulation_results_ul_k10}
    \end{minipage}
    \caption{Theoretical BEP and simulated BER for uplink ND-NOMA versus $\delta = {\sigma_{2,l}^2}/{\sigma_w^2}$, with Rician $K$-factor values of (a) $K = 5$ and (b) $K = 10$ dB}
    \label{fig:ALL_Uplink}
\end{figure*}

To use the same analysis of Section III.C, we again resort to the strong law of large numbers by assuming $s_{y_2}^2 \approx \frac{1}{N-1} \sum\nolimits_{n=1}^{N} \left| y^n_2 - h_2 m_{1,i} \right|^2$. Expressing, this new $s_{y_2}^2$ in the quadratic form of $2N$ real Gaussian random variables as $s_{y_2}^2=\mathbf{y}^{\text{T}} \Lambda\mathbf{y}$, where $\mathbf{y}$ and $\Lambda$ are as defined in \eqref{eq:quadratic}, we obtain the same banded covariance matrix of \eqref{eq:15} with the following updated parameters:
\begin{align}
\sigma_R^2&=\mathrm{VAR}[y_{2,R}^n]= h_{2,R}^2 \sigma_{2,k}^2 + \sigma_w^2 /2 \nonumber \\
\sigma_I^2&=\mathrm{VAR}[y_{2,I}^n]= h_{2,I}^2 \sigma_{2,k}^2 + \sigma_w^2 /2 \nonumber \\
c&=\mathrm{COV}( \left( y^n_{2,R}-h_{2,R} m_{1,i}  \right),\left( y^n_{2,I}-h_{2,I} m_{1,i}  \right) ) \nonumber \\
&=h_{2,R} h_{2,I} \sigma_{2,k}^2.
\label{eq:27}
\end{align}
Considering CLT and substituting the new values of \eqref{eq:27} first in \eqref{eq:16}, then updating \eqref{eq:17}-\eqref{eq:19} accordingly, conditional BEP is obtained.

As mentioned in the uplink scheme, the detection methods employed maintain the same computational complexity, avoiding the intensive iterative interference cancellation and numerical computations required in traditional PAM demodulation with SIC.

The following steps outline the key aspects of the downlink system model, summarizing the transmission and detection processes:

\begin{itemize}
    \item \textit{Step 1}:
    The BS transmits a Gaussian noise signal where the mean and variance encode U$_1$'s and U$_2$'s bits, respectively.

    \item \textit{Step 2}:
    U$_1$ estimates its bit by computing the mean of received samples using a minimum distance detector. 

    \item \textit{Step 3}:
    U$_2$ estimates its bit by computing the variance of received samples using a threshold-based detector. 
\end{itemize}

\section{Numerical Results}

\begin{table}[t!]
\centering
\caption{Simulation Parameters}
\renewcommand{\arraystretch}{1.2} 
\begin{tabular}{|p{2.5cm}|p{2.5cm}|} 
\hline
\textbf{Parameter} & \textbf{Value} \\ \hline \hline
$K_\text{{dB}}$  & 5, 10 \\ \hline
$N$       & 50, 100 \\ \hline
$\delta_\text{{dB}}$ & [-40, 5] \\ \hline
$\alpha$  & 10 \\ \hline
$P_\text{{dBm}}$ & 30 \\ \hline
$\beta$   & 1/100 \\ \hline
\end{tabular}
\label{tab:ber_results}
\end{table}

In this section, we analyze the bit error rate (BER) performance of the proposed ND-NOMA system in both uplink and downlink scenarios. In our analysis, the channel coefficients are modeled as if their envelopes are Rician distributed, $h_{1,R},h_{1,I},h_{2,R},h_{2,I} \sim \mathcal{N}\left(\sqrt{\frac{K}{2(1+K)}},\frac{1}{2(1+K)}\right)$, where $K$ being the Rician factor. This choice of distribution reflects the presence of a line-of-sight component along with multipath effects, capturing the realistic propagation conditions for the wireless channels considered in our study. In order to ensure accurate estimation of noise statistics, the number of samples per symbol duration, $N$, should be selected as $N \geq 50$, which corresponds to a sufficiently high sampling rate at the receiver. In addition to Rician fading, we also assess system performance under Rayleigh fading channels by setting $K=0$, which corresponds to environments where no direct line-of-sight component is present. Moreover, we employed the Monte Carlo integration method, as described in Appendix A, to compute the unconditional BEP for uplink and downlink scenarios so as to verify our computer simulation results. The simulation parameters are outlined in Table \ref{tab:ber_results}.

\subsection{Uplink Scenario}
Figs. 4(a) and 4(b) present the BER performance of U$_1$ and U$_2$ in both the uplink and downlink ND-NOMA systems under Rayleigh fading conditions, corresponding to a Rician factor of $K=0$. The results are shown for two different sampling rates, $N=50$ and $N=100$. As illustrated, the simulated BER curves for both users exhibit an exact match with the theoretical BEP predictions, thereby validating the accuracy of our analytical framework. Notably, increasing the number of noise samples improves the detection performance, particularly for U$_1$, due to the enhanced estimation of the noise statistics. 

Figs. 5(a) and 5(b) illustrate the BER performance of U$_1$ and U$_2$ for $K \in \{5, 10\}$ respectively. Herein, the increase in the Rician $K$-factor improves the BER performance of the proposed ND-NOMA system. This enhancement is attributed to the stronger line-of-sight component in the Rician channel, which enhances detection performance for both U$_1$ and $U_2$ in the uplink scenario. In Fig. 5(b) where $K=10$, $U_1$ exhibits better BER performance compared to U$_2$, since U$_1$'s detection mechanism leverages the mean of the transmitted Gaussian samples, which is more resilient to noise variations compared to U$_2$'s variance-based detection in our scenario.

The impact of $\delta$ on BER is also critical in the uplink scenario. In Figs. 4, 5, and 6, increasing $\delta$, which represents the higher ratio of useful to disruptive noise variances, results in improved BER performance as depicted in Figs. 4(a) and 4(b). However, in Fig. 5(b), after the $\delta$ value of $-5$, BER saturates where $N=50$ for U$_1$, since there is an interference between the users. This trend is evident in Figs. 6(a) and 6(b), where higher $\delta$  values correlate with improved BER performance. The ability of ND-NOMA to maintain low BER even in scenarios with significant noise variance is demonstrated, indicating its effectiveness in environments with varying noise conditions. It is worth noting that the users of ND-NOMA are not interference-free; nevertheless, its unique nature allows the BS to distinguish user signals without SIC.

Furthermore, another important parameter that significantly affects the BER performance of the proposed ND-NOMA system is the number of noise samples per bit ($N$), which stands for the number of samples that are taken to estimate the mean for U$_1$ and estimate the variance for U$_2$. As can be seen in Figs. 5(a) and 5(b), increasing $N$ boosts the performance of the system's uplink scenario. Our findings also underscore the importance of sample size $N$ in reducing BER, highlighting that the transition from $N$ being 50 to 100 allows better estimation accuracy of the Gaussian noise parameters, thus minimizing the impact of noise as well as improving overall detection performance.

\begin{figure*}[t]
    \centering
    \begin{minipage}{0.49\textwidth}
        \centering
        \includegraphics[width=\textwidth]{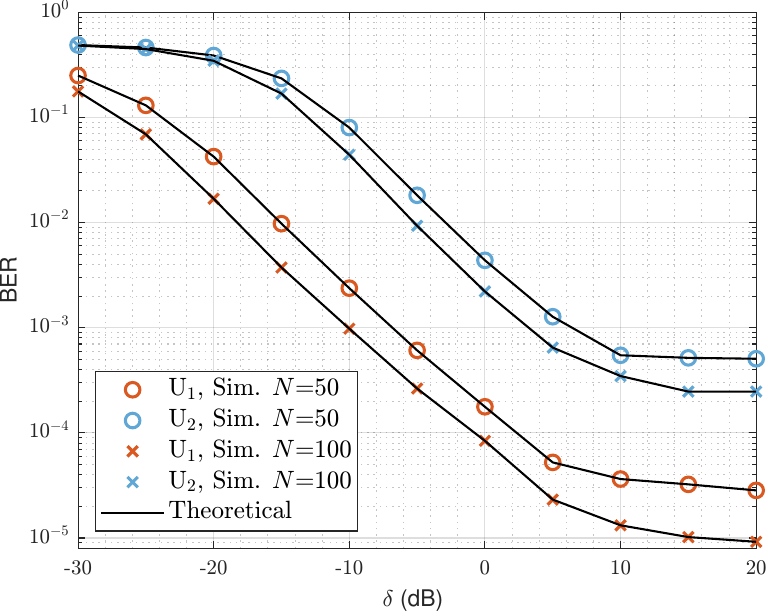}
        \\ \text{(a)}
        \label{fig:simulation_results_dl_k5}
    \end{minipage}%
    \hfill
    \begin{minipage}{0.49\textwidth}
        \centering
        \includegraphics[width=\textwidth]{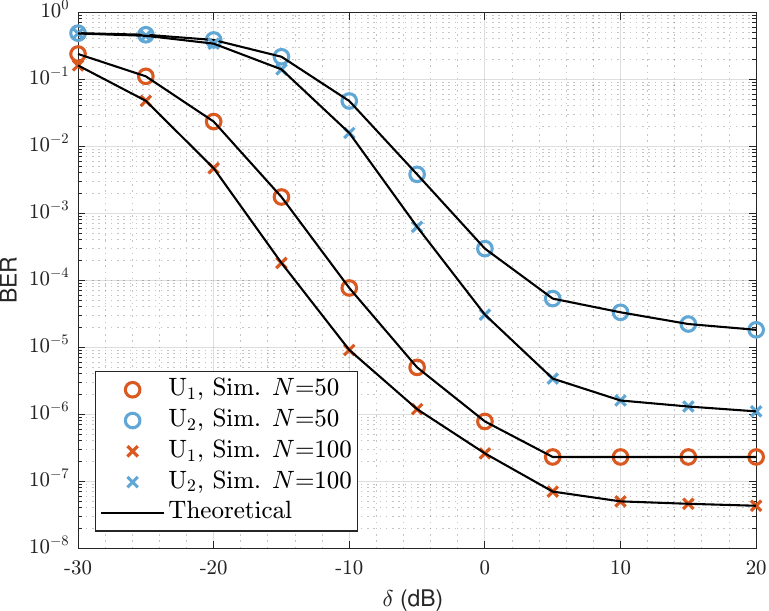}
        \\ \text{(b)}
        \label{fig:simulation_results_dl_k10}
    \end{minipage}
    \caption{Theoretical BEP and simulated BER for downlink ND-NOMA versus $\delta = {\sigma_{2,l}^2}/{\sigma_w^2}$, with Rician $K$-factor values of (a) $K = 5$ and (b) $K = 10$ dB}
    \label{fig:ALL_Downlink}
\end{figure*}

\subsection{Downlink Scenario}
Fig. 4(b) illustrates the BER of U$_1$ and U$_2$ in the downlink ND-NOMA system under Rayleigh fading ($K=0$). As in the uplink scenario, the downlink scenario's simulation results coincide with our theoretical derivations, in Figs. 6(a) and 6(b), demonstrating the system's reliability once again, even in the highly disruptive noise conditions. As expected, the increase in $\delta$ as well as $K$ enhances the BER performance of the proposed system in the downlink scenario for U$_1$ and U$_2$. Increasing $\delta$ as well as $N$ and $K$ improves the system performance. Different from the uplink scenario where $K=10$, the BER performance of the ND-NOMA system of U$_1$ does not saturate after a specific $\delta$ value since there is no interference.

We note that the impact of the sample size $N$ on the  BER in the downlink scenario is significant. As $N$ increases, the BER decreases for both U$_1$ and U$_2$. This is due to the fact that larger sample sizes allow for better averaging of the noise parameters, which are supported by a minimum distance detector for mean detection at U$_1$ and a threshold-based detector for variance detection at U$_2$, which improves the accuracy of related parameters. Consequently, the ND-NOMA system achieves lower BER with higher $N$, demonstrating its efficiency and reliability in maintaining high-quality communication in downlink scenarios.

Comparing Fig. 5(b) with Fig. 6(b), where uplink and downlink scenarios are considered, respectively. The BER performance of U$_2$ exhibits similar results for both scenarios. However, two bits are transmitted in the downlink scenario, unlike the uplink scenario, where one bit of information is conveyed at every transmission instant. Furthermore, U$_1$ BER gets saturated since there is an interference in the uplink scenario; however, in the downlink scenario, since we assume that U$_1$ signal does not interfere with U$_2$ signal, its performance does not reach saturation. Additionally, we have presented the OMA-NoiseMod scenario in the following subsection to prevent interference between users.

\begin{figure*}[t]
    \centering
    \begin{minipage}{0.49\textwidth}
        \centering
        \includegraphics[width=\textwidth]{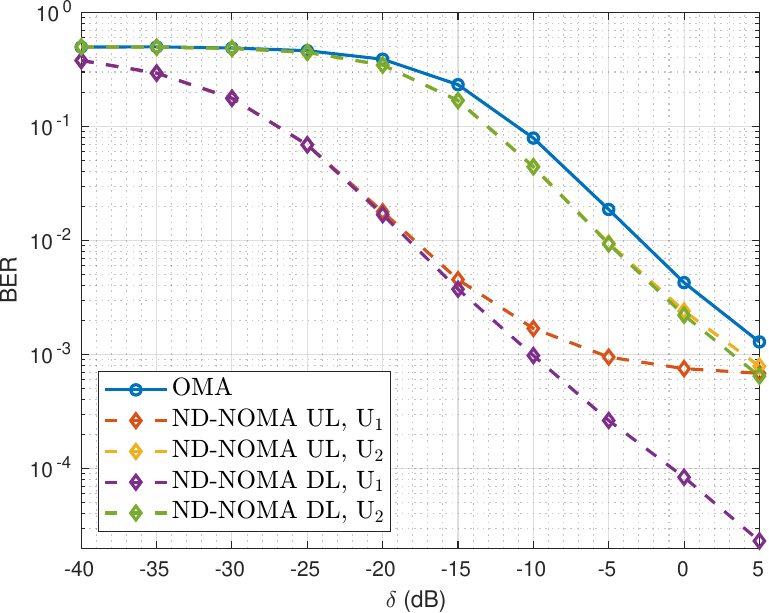}
        \\ \text{(a)}
        \label{fig:oma_5}
    \end{minipage}%
    \hfill
    \begin{minipage}{0.49\textwidth}
        \centering
        \includegraphics[width=\textwidth]{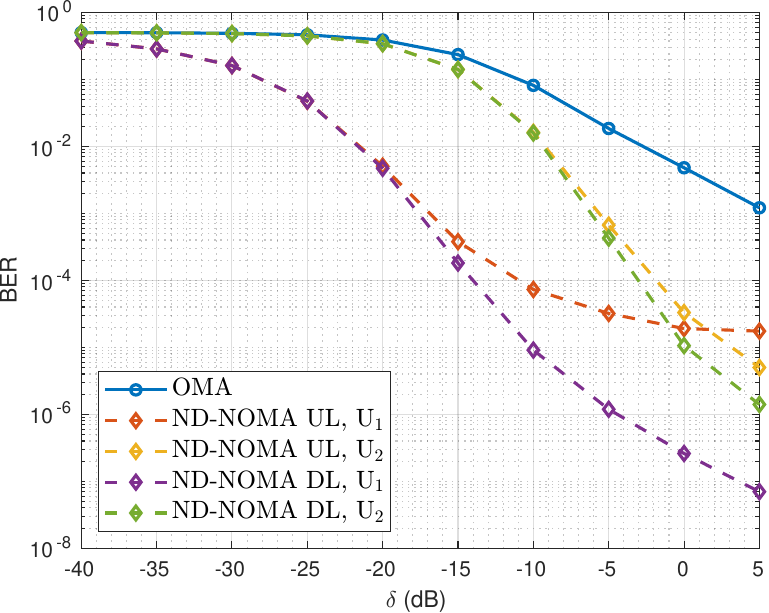}
        \\ \text{(b)}
        \label{fig:oma_10}
    \end{minipage}
    \caption{Comparative BER performance of OMA-NoiseMod and ND-NOMA (uplink and downlink) with respect to $\delta$, with Rician $K$-factor values of (a) $K = 5$ and (b) $K = 10$ dB}
    \label{fig:ALL_OMA}
\end{figure*}

\subsection{Comparison with the Benchmark Schemes}
\subsubsection{Comparison with OMA-NoiseMod Scenario}

In scenarios where two users communicate simultaneously over \( N \) noise samples in uplink/downlink, the communication can be divided into two equal parts, \( N/2 \) - \( N/2 \), allowing classic NoiseMod communication without interference. This scheme can be referred to as OMA-NoiseMod. For the evaluation of the BER of ND-NOMA under comparable modulation principles, we introduce OMA-NoiseMod as a benchmark. Unlike conventional OMA schemes employing standard amplitude and phase modulations, OMA-NoiseMod enables a fairer comparison by operating under similar noise-domain constraints. While traditional OMA inherently avoids multi-user interference and thus achieves superior BER, such a comparison would not be meaningful given the fundamentally different transmission mechanisms of traditional amplitude/phase modulation and NoiseMod schemes. It would not be a fair comparison to compare ND-NOMA schemes with the traditional (for instance, phase shift keying-based) NOMA schemes since \(\delta\) dictates the ratio between useful and disruptive noise variances, its effect on detection performance is fundamentally different.

Figs.  7(a) and  7(b) illustrate the comparative BER performance of ND-NOMA and the OMA-NoiseMod scenario for both uplink and downlink transmissions. As shown in Fig. 7(a), ND-NOMA outperforms OMA-NoiseMod in terms of BER under the same conditions, particularly as the Rician $K$-factor and $\delta$ increase. 

In the higher Rician $K$-factor scenario ($K=10$), as shown in Fig. 7(b), ND-NOMA continues to exhibit superior BER performance compared to OMA-NoiseMod. The key advantage of ND-NOMA over OMA-NoiseMod lies in its ability to manage user interference through the simultaneous transmission of data. The results clearly show that with an increased number of noise samples ($N$) and higher Rician $K$-factors, ND-NOMA achieves lower BER compared to OMA-NoiseMod. This makes ND-NOMA a more efficient and reliable communication scheme.

In the OMA-NoiseMod scenario, both U$_1$ and U$_2$ experience the same BER performance because the communication process is divided into non-overlapping time slots or frequency bands, ensuring that both users have identical conditions for transmitting and receiving data. This means there is no interference between the users, and both experience the same signal quality and noise conditions, leading to identical BER outcomes. This explains the reason why their performances are the same in the OMA-NoiseMod scheme.

\begin{figure}[t]
    \centering
    \includegraphics[width=\linewidth]{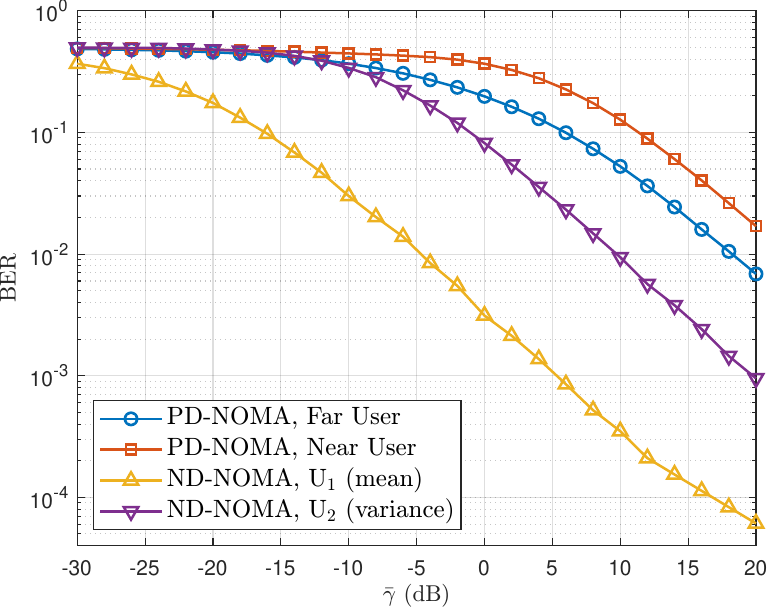}
    \caption{Comparative BER performance of the proposed ND-NOMA downlink scheme and the conventional power-domain PD-NOMA scheme under Rayleigh fading.}
    \label{fig:downlinkpdnoma}
\end{figure}

\subsubsection{Comparison with PD-NOMA}
Fig.~\ref{fig:downlinkpdnoma} shows the BER comparison of ND-NOMA and PD-NOMA over the same average SNR range, defined as $\bar{\gamma} = P_{\mathrm{tot}}/\sigma_w^2$ with $P_{\mathrm{tot}}=1$, which directly yields $\sigma_w^2 = 1/\bar{\gamma}$. This ensures that both schemes are simulated under an identical noise power for a given $\bar{\gamma}$ value, aligning with the emphasis in \cite{11124558} on the necessity of fair and consistent benchmarking between NOMA schemes. For ND-NOMA, the variance-domain parameter $\delta = \sigma_{\mathrm{useful}}^2/\sigma_w^2$ is linked to $\bar{\gamma}$ through $\sigma_{\mathrm{useful}}^2 = \delta/\bar{\gamma}$, guaranteeing that the same $\bar{\gamma}$ axis also reflects equivalent operating conditions in the variance domain. The simulations were conducted over a Rayleigh fading channel with $N=150$ noise samples per transmitted bit and $\rho_{\mathrm{far}}=0.8$, following a typical asymmetric power allocation in PD-NOMA.

The results clearly indicate that ND-NOMA achieves consistently lower BER than PD-NOMA across the examined $\bar{\gamma}$ range. This performance gain arises from ND-NOMA’s noise-domain encoding, which inherently avoids the need for SIC and thereby eliminates error propagation. Conversely, PD-NOMA’s dependence on accurate interference cancellation renders it more susceptible to fading and additive noise, particularly at lower $\bar{\gamma}$ values. By adopting a common $\bar{\gamma}$ definition and normalizing both schemes to the same total transmit power, the observed BER advantage of ND-NOMA can be attributed solely to its encoding strategy rather than differences in power scaling or SNR interpretation.

\section{Conclusion}
In this paper, we have introduced ND-NOMA, an innovative communication scheme that leverages noise variance and mean to transmit data. ND-NOMA reduces power consumption and complexity, making it a promising solution for IoT networks. Through theoretical analysis and simulations, we have demonstrated that ND-NOMA achieves low bit error rates in both uplink and downlink scenarios, outperforming traditional OMA-NoiseMod in spectral efficiency and interference management. The ND-NOMA system employs a minimum distance detector for mean detection and a threshold-based detector for variance detection, ensuring reliable performance even in low-power environments. In future works, new noise-like multiple-accessing techniques will be explored further to enhance physical layer security in addition to their communication benefits. Exploration of non-coherent ND-NOMA solutions is also an interesting research direction to further simplify the receiver architectures. To further extend the scheme to support more users, manipulation of the additional noise features, such as correlation, can be implemented, for example, encoding the third user's information through the correlation between the Gaussian noise samples, while the first and second users are encoded via mean and variance, respectively. This can be a promising scheme. While this work primarily focuses on theoretical modeling and simulations, practical challenges such as hardware-induced noise distortions, limitations in controlling statistical noise features, and the need for precise synchronization may arise in real-world implementations, which will also be investigated in our future studies.
\appendices

\section{Monte Carlo Integration Method}
\label{sec:monte_carlo}
Monte Carlo integration is a powerful technique for numerically estimating integrals, particularly in high-dimensional spaces or when the integrand has a complex form. In this appendix, we have addressed the problem of unconditional BEP integrals mentioned in previous sections. As an initial step, we can rewrite (\ref{eq:9}) as
\begin{align}
\bar{P}_b=\!\!\int_V g(h_{1,R},h_{1,I},h_{2,R},h_{2,I}) d\hat{h}, 
\label{eq:integrand_1}
\end{align}
where 
\begin{equation}
    g(h_{1,R},h_{1,I},h_{2,R},h_{2,I})=P_b f(h_{1,R}) f(h_{1,I}) f(h_{2,R}) f(h_{2,I}) \nonumber
\end{equation}
and $d\hat{h}=d h_{1,R} d h_{1,I}d h_{2,R} d h_{2,I}$. Here, \( V\) is the whole integration volume. Then, we will utilize importance sampling to compute the integral efficiently. We choose an appropriate joint PDF \( z(h_{1,R},h_{1,I},h_{2,R},h_{2,I}) \) that matches the form of the integrand as
\begin{equation}
z(h_{1,R},h_{1,I},h_{2,R},h_{2,I}) = z(h_{1,R}) z(h_{1,I})z(h_{2,R})z(h_{2,I}),
\label{eq:joint_pdf}
\end{equation}
where \( z(\cdot) \) denotes the sampling function in the integral volume $V (z > 0, \int_V f_n \, \mathrm{d}V = 1)$. Specifically, \( z(\cdot) \) represents a PDF characterized by the same mean and variance as the PDFs of channel gains \( f(\cdot) \), as described in (\ref{eq:9}). Incorporating this expression, (\ref{eq:integrand_1}) can be reformulated as
\begin{align}
\bar{P}_b=\!\!\int_V \frac{g(h_{1,R},h_{1,I},h_{2,R},h_{2,I})z(h_{1,R},h_{1,I},h_{2,R},h_{2,I})}{z(h_{1,R},h_{1,I},h_{2,R},h_{2,I})}d\hat{h}
\label{eq:integrand_2}
\end{align}

Afterwards, we generate \( J \) random points for all integral variables from corresponding sampling functions as $\{\mathbf{h}^{(1)},\mathbf{h}^{(2)},\ldots,\mathbf{h}^{(J)}\}$ where \( \mathbf{h}^{(j)}=[h_{1,R}^{(j)}, h_{1,I}^{(j)},h_{2,R}^{(j)}, h_{2,I}^{(j)}]\) for $j=1,2,\ldots,J$. Then the importance weights are calculated as follows
\begin{equation}
w(\mathbf{h}^{(j)}) = \frac{g(h_{1,R}^{(j)},h_{1,I}^{(j)},h_{2,R}^{(j)},h_{2,I}^{(j)})z(h_{1,R}^{(j)},h_{1,I}^{(j)},h_{2,R}^{(j)},h_{2,I}^{(j)})}{z(h_{1,R}^{(j)},h_{1,I}^{(j)},h_{2,R}^{(j)},h_{2,I}^{(j)})}.
\label{eq:weights}
\end{equation}
The integral is estimated using the average of these weights as expressed below: \cite{Monte_Carlo_Integration_2020}
\begin{equation}
\Bar{P}_b = \frac{1}{J} \sum_{j=1}^{J} w(\mathbf{h}^{(j)}).
\label{eq:mc_estimation}
\end{equation}
For unconditional BEP curves, $J=10^6$ random sample points are used for all integral estimations.

\bibliographystyle{IEEEtran}
\bibliography{IEEEabrv,bib_2023}

\begin{IEEEbiography}[{\includegraphics[width=1in,height=1.25in,clip,keepaspectratio]{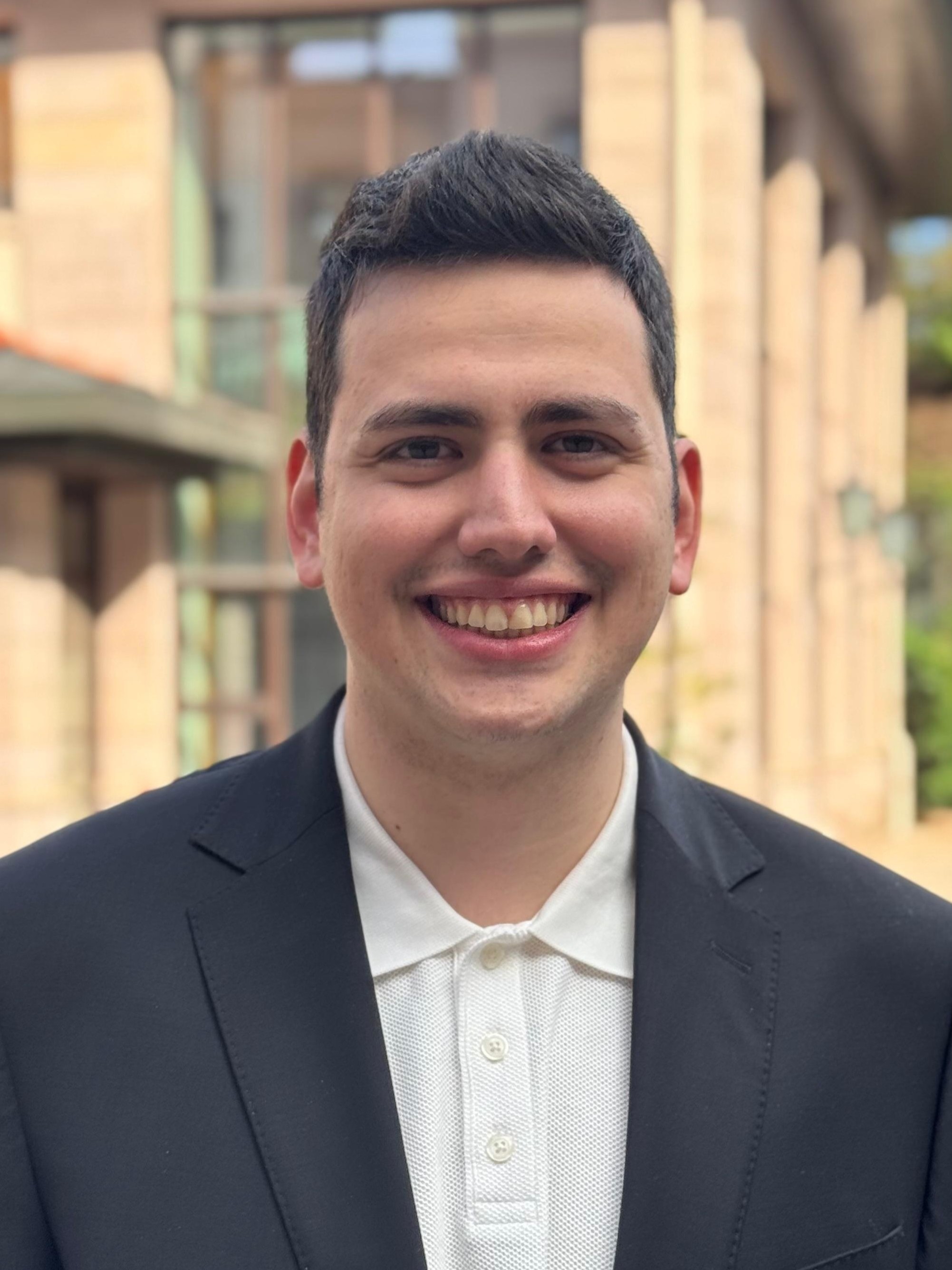}}]{ERKIN YAPICI } (Graduate Member, IEEE) received the B.Sc. and M.Sc. degrees in electrical and electronics engineering from Koç University, Istanbul, Turkey, in 2023 and 2025, respectively. He is currently pursuing his Ph.D. degree at Boğaziçi University, Istanbul. During his M.Sc. studies, he was a Research and Teaching Assistant at Koç University. 

His research interests include noise modulation, wireless energy transfer, Internet of Things, and next-generation wireless communications.
\end{IEEEbiography}

\begin{IEEEbiography}[{\includegraphics[width=1in,height=1.25in,clip,keepaspectratio]{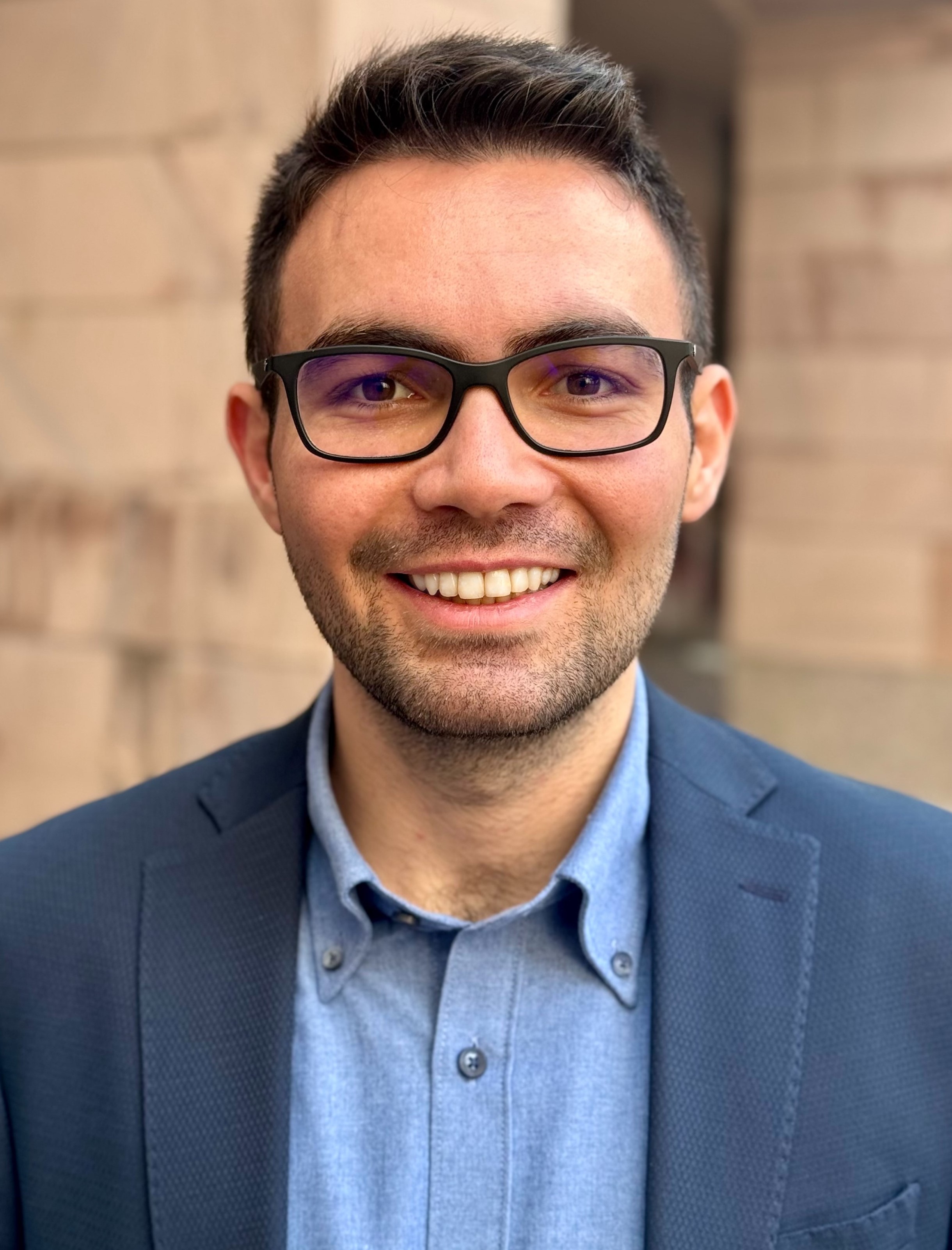}}]{YUSUF ISLAM TEK }
 (Graduate Member, IEEE) received his B.Sc. degree in Electrical and Electronics Engineering from Nuh Naci Yazgan University, Kayseri, Turkey, in 2020, and the M.Sc. degree from Koç University, Istanbul, in 2023, where he is currently pursuing his Ph.D. degree. He is currently a Research and Development Engineer at Türk Telekom. 
 
 His research interests include next-generation wireless networks, physical layer security, waveform design, noise-driven communications, signal processing for communications, and index modulation. He served as a Reviewer for several IEEE letters and journals.
\end{IEEEbiography}

\begin{IEEEbiography}[{\includegraphics[width=1in,height=1.25in,clip,keepaspectratio]{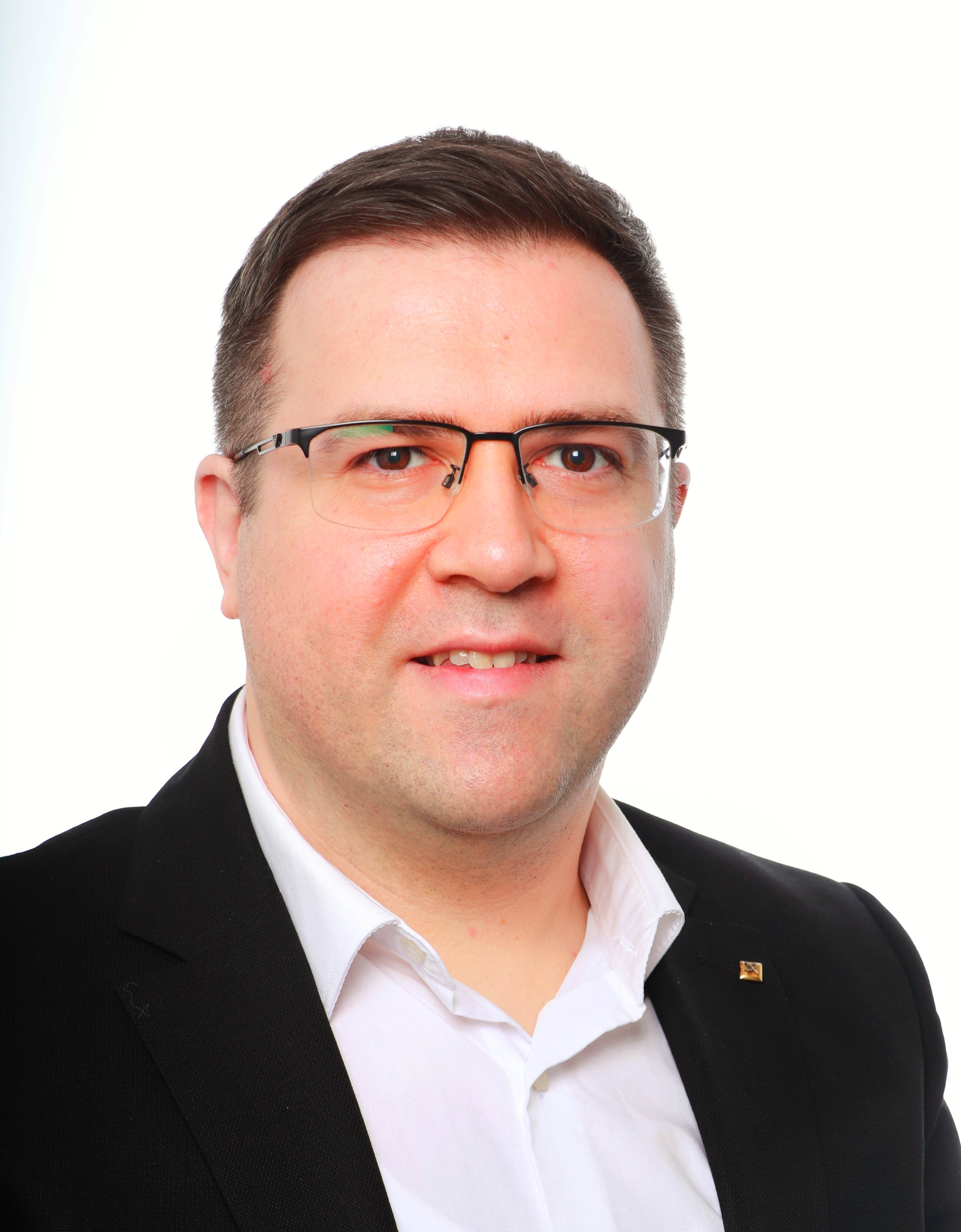}}]{ERTUGRUL BASAR } (Fellow, IEEE) Prof. Ertugrul Basar was born in Istanbul, Turkey, in 1985. He received the B.S. degree (with High Honors) from Istanbul University, Turkey, in 2007 and the M.Sc. and Ph.D. degrees from Istanbul Technical University, Turkey, in 2009 and 2013, respectively. He is a Professor of Wireless Systems at the Department of Electrical Engineering, Tampere University, Finland. Before joining Tampere University, he held positions at Koç University from 2018 to 2025 and at Istanbul Technical University from 2009 to 2018. He also had visiting positions at Princeton University, Princeton, NJ, USA, as a Visiting Research Collaborator from 2011 to 2022, and at Ruhr University Bochum, Bochum, Germany, as a Mercator Fellow in 2022. 

Prof. Basar’s primary research interests include 6G and beyond wireless communication systems, multi-antenna systems, index modulation, reconfigurable intelligent surfaces, waveform design, zero-power and thermal noise communications, software-defined radio, physical layer security, quantum key distribution systems, and signal processing/deep learning for communications. He is the author/co-author of around 200 international journal publications that have received more than 18,000 citations.
\end{IEEEbiography}
\balance
\end{document}